\documentclass[aps,prl,twocolumn,superscriptaddress,hidelinks,nobibnotes,nofootinbib]{revtex4-2}
\usepackage[T1]{fontenc}
\usepackage{graphicx}
\usepackage{physics}
\usepackage{amssymb}
\usepackage{hyperref}
\usepackage{xcolor}
\usepackage{soul}
\usepackage{bm}
\usepackage{mathtools}
\usepackage[symbol, hang]{footmisc}

\usepackage{esvect}

\hypersetup{
    colorlinks,
    linkcolor=[RGB]{46, 48, 146},
    citecolor=[RGB]{46, 48, 146},
    urlcolor=[RGB]{46, 48, 146}
}

\begin{document}

\title{Compatibility of Trapped Ions and Dielectrics at Cryogenic Temperatures}

\begin{abstract}
\noindent We study the impact of an unshielded dielectric\textemdash here, a bare optical fiber\textemdash on a $^{40}$Ca${^+}$ ion held several hundred microns away in a cryogenic surface electrode trap. We observe distance-dependent stray electric fields of up to a few kV/m due to the dielectric, which drift on average less than 10\,\% per month and can be fully compensated with reasonable voltages on the trap electrodes. We observe ion motional heating rates attributable to the dielectric of $\approx30$ quanta per second at an ion-fiber distance of 215(4)\,$\mu$m and $\approx$1.5\,MHz motional frequency. These results demonstrate the viability of using unshielded, trap-integrated dielectric objects such as miniature optical cavities or other optical elements in cryogenic surface electrode ion traps.  
\end{abstract}

\author{M. Bruff}
\thanks{These authors contributed equally to this work.}
\affiliation{National Institute of Standards and Technology, 325 Broadway, Boulder, CO 80305, USA}
\affiliation{Department of Physics, University of Colorado, Boulder, CO 80309, USA}

\author{L. Sonderhouse}
\thanks{These authors contributed equally to this work.}
\affiliation{National Institute of Standards and Technology, 325 Broadway, Boulder, CO 80305, USA}

\author{{K. N. David}}
\affiliation{National Institute of Standards and Technology, 325 Broadway, Boulder, CO 80305, USA}
\affiliation{Department of Physics, University of Colorado, Boulder, CO 80309, USA}

\author{J. Stuart}
\thanks{Current address: IonQ, 4505 Campus Drive, College Park, MD 20740}
\affiliation{National Institute of Standards and Technology, 325 Broadway, Boulder, CO 80305, USA}

\author{D. H. Slichter}
\affiliation{National Institute of Standards and Technology, 325 Broadway, Boulder, CO 80305, USA}

\author{D. Leibfried}
\affiliation{National Institute of Standards and Technology, 325 Broadway, Boulder, CO 80305, USA}
\maketitle

Trapped ions are a leading platform for scalable quantum computation~\cite{moses_race_2023, chen_benchmarking_2024, loschnauer_scalable_2024} and realize some of the most accurate optical atomic clocks~\cite{marshall_highstability_2025, huntemann_clock_2016, hausser_clock_2025}. Because they can be entangled with optical photons serving as ``flying qubits,'' trapped ions are also promising candidates for the ``stationary qubit'' nodes of a quantum network~\cite{duan_longdistance_2001, duan_scalable_2004, briegel_quantumrepeaters_1998}. Their long coherence times~\cite{wang_singleion_2021} and high-fidelity quantum logic operations~\cite{loschnauer_scalable_2024, smith_singlequbit_2025, clark_highfidelity_2021} are advantageous for entanglement distillation and entanglement swapping protocols~\cite{bennett_purification_1996,briegel_quantumrepeaters_1998}, and they can be directly used in distributed quantum computing systems~\cite{Monroe_largescale_2014} and networks of entangled atomic clocks~\cite{nichol_elementary_2022}.

Bell states of remotely entangled trapped ions have been created with fidelities up to 97\,\%, and fidelities above 99.9\,\% appear feasible~\cite{saha_highfidelity_2024}. However, photon collection efficiencies have limited entanglement rates to 250\,s$^{-1}$ between systems separated by a few meters~\cite{oreilly_fastphoton_2024}, far below the 10\,kHz to 100\,kHz rate at which local quantum gates are typically performed~\cite{moses_race_2023, chen_benchmarking_2024, loschnauer_scalable_2024}. Reference~\cite{oreilly_fastphoton_2024} used a high-numerical-aperture (NA) lens to collect photons emitted from the ion, with a net collection efficiency up to 10\,\% after coupling the light into a single-mode fiber~\cite{carter_iontrap_2024}. High-finesse optical cavities with a small ($\lesssim10\,\mu$m) mode waist are an attractive alternative that can provide near-unity collection efficiency~\cite{schupp_interface_2021}. However, realizing the requisite high finesse becomes increasingly difficult for longer cavities. Shorter cavities with smaller micromachined mirrors, potentially on the ends of optical fibers, thus could be advantageous~\cite{hunger_fiber_2010, brekenfeld_quantum_2020, kobel_deterministic_2021, Pfeifer2022, riedel_deterministic_2017, deshmukh_dielectric_2023, gao_optimization_2023, kay_carbon_2024, grinkemeyer_errordetected_2025, gao_profile_2025}.

Trapping an ion in a shorter cavity brings it into closer proximity with the dielectric cavity mirrors and substrates, which as insulators can host static and oscillating charges on their surfaces and in the bulk. This can present a challenge as ions are highly sensitive to the resulting electric fields~\cite{Turchette2000, huber_trappedion_2010, narayanan_electricfield_2011, brownnutt_iontrap_2015}. The question of system compatibility with nearby dielectrics is relevant not only to trapped ions, including trap-integrated optical cavities~\cite{Sterk2012, vanrynbach_integrated_2016, teller_integrating_2023, takahashi_strongcoupling_2020, kobel_deterministic_2021, Kassa2025} and trap-integrated laser delivery and fluorescence collection optics~\cite{brady_integration_2011, kim_surfaceelectrode_2011, VanDevender_efficient_2010, jechow_quantum_2011,  takahashi_integrated_2013, Bado2015,mehta_inegrated_optics_2016,niffenegger_multiwav_2020, mehta_multiion_2020, Day2021,Knollmann2024, quantinuum_integrated_2023,fang_cavity_2025}, but also to any electric-field sensitive system such as neutral Rydberg atoms~\cite{Bernien2017, gu_hybrid_2024} or Casimir force measurements~\cite{garrett_measuring_2020}.
 Conductive coatings such as indium tin oxide (ITO) can reduce charging, but also introduce optical loss that prevents their use in high-finesse optical cavities~\cite{eltony_transparent_2013, chen_design_2025}. Previous room-temperature experiments have measured large stray electric fields near unshielded dielectrics that drift over hours and displace the ion by hundreds of $\mu$m~\cite{ong_probingsurface_2020}. These stray fields are caused by charges that accumulate on the surface of the dielectric. Thermally activated fluctuators inside the dielectric bulk radiate oscillating electric fields that cause ion motional heating; the heating rate of 70,000\,quanta/s at a 1.6\,MHz mode frequency for a 250\,$\mu$m ion-dielectric separation measured in Reference~\cite{teller_heating_2021} is high enough to impede ground state cooling, reduce coupling to the optical cavity~\cite{roos_controlling_2000, schupp_interface_2021}, and impact the fidelity of ion-ion remote entanglement~\cite{saha_highfidelity_2024, apolin_recoilinduced_2025, kikura_recoil_2025} and local two-qubit gates~\cite{sorensen_quantum_1999, sorensen_entanglement_2000, Milburn2000}. While these studies were carried out at room temperature, cryogenically operated ion traps are becoming increasingly common due to a variety of advantages~\cite{moses_race_2023, Pagano2018, dubielzig_ultralowvibration_2021, brownnutt_iontrap_2015, susanna_scalable_2020, deslauriers_scaling_2006}, and may exhibit different behavior for trap-integrated dielectrics.
 
Here, we report measurements of trapped ion behavior in the presence of a nearby unshielded dielectric at cryogenic temperatures. We use a linear surface-electrode ion trap operated at $\approx\,6.5$\,K with a bare optical fiber placed on top of the trap chip and aligned along the trap axis. We determine the three-dimensional stray electric field $\vec{E}$ at various ion-fiber distances $d$, compensate it using voltages applied to the trap electrodes, and track an overall slow decrease over several months in the electric field magnitude along the trap axis. We fit the data with a simple model of fields from charges accumulated on the fiber surface, yielding equivalent stray charge densities similar to those reported in Reference~\cite{ong_probingsurface_2020}. We also measure motional heating rates both along and perpendicular to the ion-fiber axis in a plane parallel to the electrodes for different $d$. The equivalent electric field noise power spectral density $S_E(\omega)$ is ${\sim10^3}$ times lower than that reported in Reference~\cite{teller_heating_2021} for equivalent distances and mode orientations, due to a combination of the cryogenic temperature and electric field shielding from the trap electrodes themselves. Modeling the fiber as a body with uniform relative permittivity $\mathrm{ \epsilon_r}$ and dielectric loss tangent $\mathrm{\tan \delta}$, we extract the product $\mathrm{\epsilon_r \tan \delta} = 0.0050(7)$.

\begin{figure}
\includegraphics[width=8.4cm]{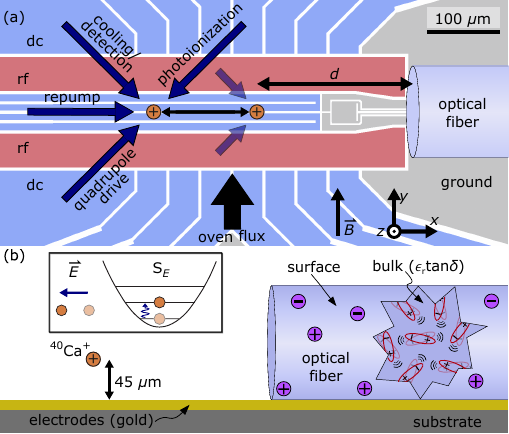}
\caption{Schematic of apparatus and relevant mechanisms. (a) Top view of the surface-electrode ion trap with attached bare optical fiber. An ion (orange circle) is loaded $\approx 325\,\mu$m from the optical fiber and transported to various ion-fiber distances $d$ to quantify stray fields and heating rates. Directions of laser beams, neutral Ca flux, and applied magnetic field $\vec{B}$ are indicated with arrows, and laser beam orientations at the transported ion position are indicated with shorter arrows. (b) Stray charges on the optical fiber surface produce a static electric field $\vec{E}$, while thermal fluctuators inside the bulk (cutaway view) radiate electric field noise with power spectral density $S_E(\omega)$ that heats ion motion.}
\label{fig:1}
\end{figure}

We perform these experiments using a single $^{40}\textrm{Ca}^+$ ion loaded into a linear surface electrode trap with a bare 125 $\mu$m-diameter optical fiber (780HP) attached to the surface, in line with the trap axis as shown in Figure~\ref{fig:1}(a)~\cite{supplement}. The trap and optical fiber are housed inside an ultrahigh vacuum chamber and are cooled to $\approx\,6.5$\,K using a closed-cycle gaseous helium flow cryostat. The rf electrodes (red) provide confinement radially (in the $y$-$z$ plane) along the rf null line, which is along the $x$ axis $\approx$\,45\,$\mu$m above the trap surface. Voltages applied to the segmented dc electrodes (blue) confine the ion axially (along $x$) and tilt the radial modes $\approx\,10^{\circ}$ from the $z$ and $y$ axes to enable efficient Doppler cooling of all modes. We refer to these as the ``out-of-plane'' and ``in-plane'' radial modes, respectively. Typical motional frequencies are 4\,MHz to 5\,MHz in the radial directions and ${\approx1.5}$\,MHz in the axial direction. A magnetic field of $\vec{B} \approx$\,0.4\,mT along $y$ defines the quantization axis.

To study the ion behavior near the optical fiber, a $^{40}\textrm{Ca}^+$ ion is first loaded at $d \approx 325\,\mu$m by photoionizing neutral calcium emitted from a thermal oven using 375 nm and 423 nm laser light~\cite{gulde_simple_2001}. Loading far from the fiber reduces stray charge accumulation on the fiber surface. Voltages on the dc electrodes are then ramped to transport the ion along the rf null line, as close as $d=215(4)\,\mu$m. Doppler cooling and fluorescence detection are performed using 397\,nm laser light to drive the $4^2S_{1/2} \leftrightarrow 4^2P_{1/2}$ transition along with 866\,nm laser light to repump the population out of the metastable $3^2D_{3/2}$ state~\cite{roos_controlling_2000}. Ion fluorescence is collected with an objective (NA\,=\,0.2) and detected with either an electron-multiplied charge-coupled device (EMCCD) camera or a photomultiplier tube (PMT). State preparation, sideband cooling, and coherent operations are performed using the quadrupole $4^2S_{1/2} \leftrightarrow 3^2D_{5/2}$ transition, which is driven using a narrow 729\,nm laser locked to an ultrastable cavity. We prepare the state $4^2S_{1/2}\ket{m_J=-1/2}\equiv\ket{\downarrow}$ by optical pumping with the 729\,nm laser tuned to the $4^2S_{1/2}\ket{m_J=1/2} \leftrightarrow 3^2D_{5/2}\ket{m_J=-3/2}$ transition frequency and an 854\,nm repump laser. Coherent operations are then driven between $\ket{\downarrow} \leftrightarrow 3^2D_{5/2}\ket{m_J=-5/2}\equiv\ket{\uparrow}$. To quantify how the bare optical fiber modifies trapped-ion performance, we carry out a reference measurement in an identical trap without an optical fiber. The heating rates and stray electric fields observed are consistent with prior reports on this trap design \cite{susanna_scalable_2020, todaro_statereadout_2021}. Collectively we refer to these measurements as the “fiber-free reference”.

\begin{figure*}[]
\includegraphics[width=17.4cm]{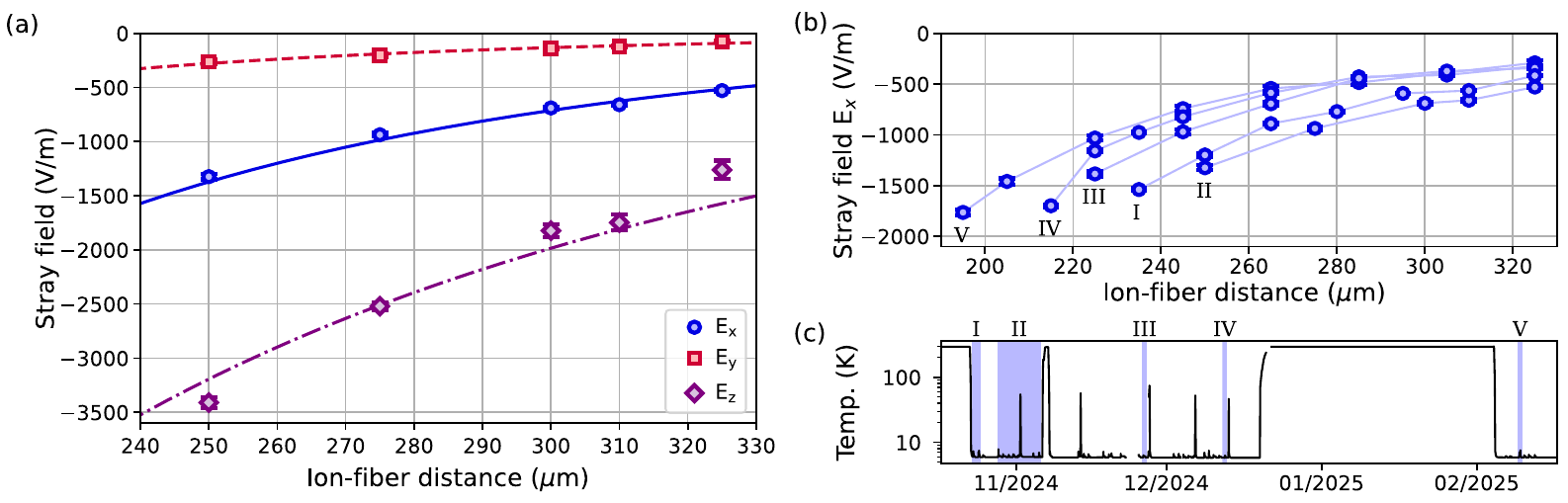}
\caption{Distance dependence of the stray electric field $\vec{E}$. (a) Stray field data in the $x$ (blue circles), $y$ (red squares), and $z$ (purple diamonds) directions. The fitted $E_x$, $E_y$, and $E_z$ are shown as lines in solid blue, dashed red, and dot-dashed purple, respectively. (b) $E_x$ from five sequential data runs [labeled I-V, where II is the data run in (a)], with connecting lines to guide the eye, showing a slow relaxation over several months. Some error bars are smaller than the plot markers. (c) Temperature $T$ of the trap and optical fiber over the time period of the data runs shown in (b), including both full and partial cycles between $\approx$\,6\,K and 300\,K. All error bars represent 68\,\% confidence intervals.}
\label{fig:2}
\end{figure*}

We use $\vec{E}$ and motional heating rates $\dot{\bar{n}}$ as a probe of static and dynamic properties of the dielectric. Figure~\ref{fig:1}(b) depicts the mechanisms we consider in our model schematically. Stray charges can accumulate on the fiber surface in several ways. Electrons can be photoelectrically dissociated from the fiber surface by stray 375 nm, 397 nm, or 423 nm laser light, leaving a net positive surface charge. Additionally, electrons dissociated from the trap electrodes or produced from calcium atoms during photoionization, along with calcium ions that we fail to trap or other charged particles, could adsorb to the fiber surface.
These charges are largely immobile on the insulating dielectric, particularly at cryogenic temperatures, and will contribute to $\vec{E}$ at the ion position. A single elementary charge on the end of the optical fiber can produce $|\vec{E}|$ as large as $\approx$\,10 mV/m at the ion position for $d = 250\,\mu$m.

Inside the bulk of the fiber, an ensemble of thermally excited fluctuators with electric dipole moments radiate a time-dependent electric field that can excite the motion of the trapped ion. These electric field fluctuations can be described by their power spectral density at the ion position $S_E(\omega, \hat{\beta})$ (in V$^2$\,m$^{-2}$\,Hz$^{-1}$) at angular frequency $\omega$ along the unit vector $\hat{\beta}$. Since direct calculation of $S_E(\omega, \hat{\beta})$ can be difficult, we use the fluctuation dissipation theorem to determine $S_E(\omega, \hat{\beta})$ by instead calculating the change in the electric field from the ion $\vec{E}_\mathrm{ion}(\vec{r}, \vec{r}\,')$ seen at position $\vec{r}\,'$ inside the dielectric bulk due to the ion being displaced along $\hat{\beta}$ from its equilibrium position $\vec{r}$~\cite{teller_heating_2021,kumph_electric-field_2016}: 
\begin{equation}
    S_E(\omega, \hat{\beta}) = \frac{4k_BT}{q^2\omega} \epsilon_0\epsilon_r\tan\delta \int_V \sum_{i}\left|\sum_{j}\frac{\partial E_{\mathrm{ion},i}}{\partial r_j} {\beta}_j\right|^2d^3\vec{r}\,'\,.
\label{eq:S_E}
\end{equation}
The integral is taken over the volume $V$ of the dielectric bulk and $i$ and $j$ index over the three spatial dimensions. Here, $k_B$ is the Boltzmann constant, $T$ is the temperature of the dielectric, $q$ is the ion charge, $\epsilon_0$ is the vacuum permittivity, and $\epsilon_r\tan\delta$ is the imaginary part of the complex relative permittivity of the dielectric. This power spectral density is related to the motional heating rate $\dot{\bar{n}}$ for a mode along axis $\hat{\beta}$ at angular frequency $\omega$ according to~\cite{Turchette2000, brownnutt_iontrap_2015}
\begin{equation}
    \dot{\bar{n}}=\frac{q^2}{4m\hbar\omega}S_E(\omega, \hat{\beta})\,,
\label{eq:nbar}
\end{equation}
where $m$ is the ion mass and $\hbar$ is the reduced Planck's constant. Equations (\ref{eq:S_E}) and (\ref{eq:nbar}) connect the measured heating rates to the intrinsic dielectric properties given the trap and dielectric geometry.

In Figure~\ref{fig:2}(a), the three components of $\vec{E}$ are determined at various ion-fiber distances by adding compensation voltages to the applied voltages on the dc electrodes so that the ion's equilibrium position is at the rf null in the $y$-$z$ plane and the unperturbed dc well minimum in $x$~\cite{supplement}. Data for a single ion-fiber distance were taken on the same day, while the full set of measurements was completed within one week. The stray fields are small enough that we can compensate them fully in all three directions over the full range of ion-fiber distances shown without exceeding $ \pm10 \textrm{ V}$ applied to the dc electrodes. The position of the optical fiber on the trap influences the magnitude of $\vec{E}$. $E_y$ is small relative to the other components due to the approximate $y$-symmetry of the optical fiber about the rf null line, while $E_z$ is the largest component because the field lines from the fiber terminate in the $z$ direction at the trap surface, normal to the conducting trap electrodes.

We can make a simple model for the origin of $\vec{E}$ by assuming a stray charge distribution on the fiber with uniform values of surface charge on the end facet, the side wall with $y>0$, and the side wall with $y<0$, where $y=0$ is along the rf null line~\cite{supplement}. We use electrostatic finite element simulations to calculate the resulting field strengths. We also assume an additional unknown uniform ``offset'' field $\vec{E}_0$ in three dimensions to account for the nonzero stray field typically seen in surface electrode traps. Fitting to the experimental data yields charge densities $\mathrm{\sigma_{f} = 9.5(6)\,e/\mu m^2}$, $\mathrm{\sigma_{+\hat{y}} = 47(9)\,e/\mu m^2}$, and $\mathrm{\sigma_{-\hat{y}} = -28(9) \,e/\mu m^2}$ for the three regions, respectively, where $e$ is the elementary charge, while the offset field from the trap is ${\vec{E}_0 = (190 \pm 60, 20 \pm 50, 620 \pm 210)}$\,V/m. The offset field is consistent with the measured stray fields from the fiber-free reference. Despite reduced charge mobility at lower temperatures, the stray charge densities on the optical fiber are of the same order of magnitude as those imputed from room temperature ion trap experiments in Reference~\cite{ong_probingsurface_2020}. This may be due to reduced charging mechanisms in our setup, such as loading ions far from the optical fiber, or due to the presence of other nearby cold surfaces to which stray charge can adhere.

We repeat the $E_x$ measurements over multiple months, as shown in Figure~\ref{fig:2}(b), and observe a slow relaxation with average drift rate <10\,\% per month, even over multiple loading attempts and temperature cycles as shown in Figure~\ref{fig:2}(c). The long-term drift of $E_y$ and $E_z$ is similar to that of $E_x$ \cite{supplement}. We observe nonmonotonic drift around data run II [Figure \ref{fig:2}(a)], which used a higher duty cycle of 397\,nm laser exposure and a higher frequency of ion loading attempts, both of which may have strengthened the stray field. The increased stability of the stray fields compared to those reported in Reference~\cite{ong_probingsurface_2020} could be due to reduced charge mobility at cryogenic temperatures.

\begin{figure}[t!]
\includegraphics[width=8.4cm]{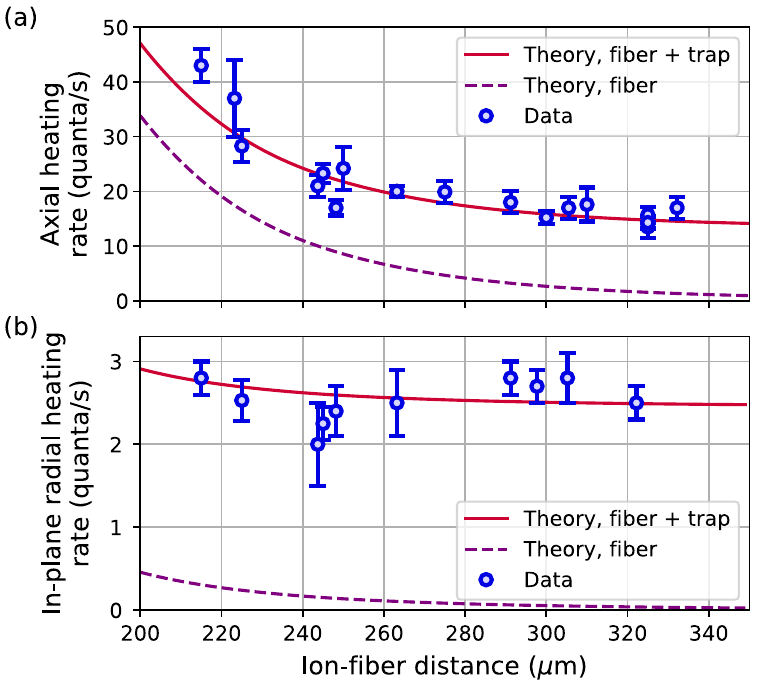}
\caption{Heating rates versus ion-fiber distance in the (a) axial and (b) in-plane radial directions. The axial data are fit with a model based on thermal fluctuations inside the optical fiber bulk plus a constant heating rate from the trap (red line). The radial data are fitted only to determine the constant heating rate from the trap, with the fiber contribution calculated using the dielectric properties from the axial fit. The axial [radial] data are taken with $\omega/2\pi = 1.45(5)$\,MHz [$4.5(6)$\,MHz], where the error bars represent 68\,\% confidence intervals. The fitted heating rate from the optical fiber alone is shown in both panels (purple, dashed).}
\label{fig3}
\end{figure}

We turn now to probing the time-dependent electric field noise from the fiber with the ion's motion. Previous room temperature experiments that trapped ions near an optical fiber found that motional heating from the dielectric was too large to allow ground state cooling~\cite{teller_heating_2021, chen_design_2025}. We are able to sideband cool the axial and in-plane radial modes to $\bar{n}<0.1$ for all ion-fiber distances shown in Figure~\ref{fig3}. Ground state cooling of the out-of-plane radial mode was limited to $\bar{n}<0.5$ by the small projection of the quadrupole drive laser beam. The motional heating rates were determined using sideband thermometry~\cite{Monroe1995a, Turchette2000,leibfried_quantum_2003}, and were taken after radial field compensation and at consistent mode frequencies of $\omega/2\pi = 1.45(5)$\,MHz (axial) and $4.5(6)$\,MHz (in-plane radial). We measure axial heating rates $<$50\,quanta/s at $d=$215(4)\,$\mu$m. We observe higher heating rates when the radial stray fields are not well-compensated, likely due to technical noise on the trap rf drive. The minimum $d$ in Figure~\ref{fig3} is constrained by the compensation limit of $\approx$\,4\,kV/m given the dc-voltage limits of $\pm 10$\,V. 

The heating rate due to the dielectric is modeled via the fluctuation-dissipation mechanism, as described in Equations~(\ref{eq:S_E}) and~(\ref{eq:nbar}). Following the approach in Reference~\cite{teller_heating_2021}, we compute the integral numerically using finite-element analysis software, using the exact surface trap geometry and optical fiber position with all electrodes grounded. Operating at 6.5\,K results in a roughly 50-fold reduction in $S_E(\omega)$ relative to room temperature. Most field lines that emanate from the dielectric terminate on the conducting trap surface due to the trap's geometry relative to the fiber and ion, reducing $S_E$ by an additional factor of $\approx\,4$\,(10) at $d=200\,\mu$m (350\,$\mu$m) when we include the trap electrodes in our simulations~\cite{supplement}.

Since to the best of our knowledge there are no data on the cryogenic dielectric properties of optical fiber at megahertz frequencies in the literature, we fit the axial heating rate data with $\mathrm{\epsilon_r\tan{\delta}}$ as a free parameter. We also include a constant offset $\mathrm{\dot{\bar{n}}_{0,a}}$ to account for the heating rate from the surface trap. From this fit, we extract $\mathrm{\epsilon_r \tan{\delta} =  0.0050(7)}$. Assuming $\mathrm{\epsilon_r} = 3.9$, the room-temperature value~\cite{jain_thinfilm_2004}, we determine a loss tangent of $\mathrm{\tan{\delta}} = 1.3(2) \times 10^{-3}$. Both the measured heating rate at $d = 330\,\mu$m and the fitted value $\mathrm{\dot{\bar{n}}_{0,a}} = 13(1)$\,quanta/s are consistent with the measured heating rates from the fiber-free reference, indicating negligible contributions from the optical fiber.

Due to the symmetry of the optical fiber with respect to the ion, thermal fluctuations in the optical fiber contribute minimally to the heating rate of the in-plane radial mode. As a result, we model the radial heating rate data by adding the dielectric heating, calculated without free parameters using the dielectric properties from the axial heating rate fit, to a constant trap background $\mathrm{\dot{\bar{n}}_{0,r}}$ that is the sole free parameter in the fit. We fit $\mathrm{\dot{\bar{n}}_{0,r}} = 2.5(1)$\,quanta/s. Although other heating mechanisms, such as technical noise or surface effects on the optical fiber, may contribute, our data are largely consistent with the fluctuation-dissipation model.

\begin{figure}
\includegraphics[width=8.4cm]{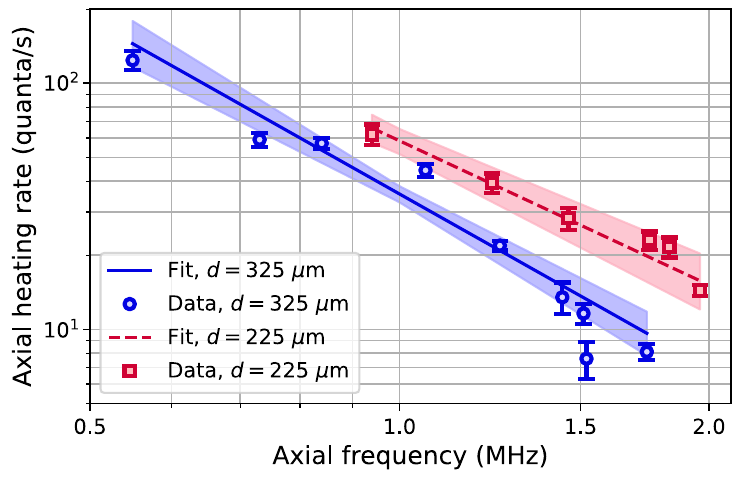}
\caption{Frequency dependence of axial heating rate at an ion position far from the fiber ($d=325(4)\,\mu$m, blue circles) and near the fiber ($d=225(4)\,\mu$m, red squares) with corresponding power law fits to the data (blue solid and red dashed lines) and 68\% confidence intervals (shaded regions).}
\label{fig4}
\end{figure}

We also study the frequency dependence of the axial heating rate at two fixed positions: one near the fiber with $d=$ 225(4)\,$\mu$m and another far from the fiber at $d=$ 325(4)\,$\mu$m. Far from the fiber, we fit the frequency scaling to a power law, $\dot{\bar{n}}_\mathrm{far}=B_\mathrm{far}\omega^{\alpha_\mathrm{far}}$ with $\alpha_\textrm{far}=-2.5(2)$. In light of the asymptotic behavior of the axial heating rate at larger $d$ [Figure \ref{fig3}a], we assume the frequency scaling far from the fiber is that of the trap. This frequency scaling is consistent with the typical frequency dependence of heating rates observed in surface-electrode ion traps~\cite{brownnutt_iontrap_2015}. 

Near the fiber, we treat the heating rate as a sum of effects from the trap and fiber, each with independent power law frequency dependence:
\begin{equation}
    \dot{\bar{n}}_\textrm{near}(\omega)=B_\textrm{trap}\omega^{\alpha_\textrm{trap}}+B_\textrm{fib}\omega^{\alpha_\textrm{fib}}.
\end{equation}
We can choose to use the values extracted far from the fiber for the trap parameters, $B_\mathrm{trap} = B_\mathrm{far}$ and $\alpha_\mathrm{trap} = \alpha_\mathrm{far}$, yielding $\alpha_\textrm{fib}=-1.5(5)$.
However, since the magnitude and frequency dependence of the trap heating rate may be ion-position dependent, we can alternatively fit the frequency dependence of heating rates near the fiber to a single power law, $\dot{\bar{n}}_\mathrm{near}=B_\mathrm{near}\omega^{\alpha_\mathrm{near}}$. This gives $\alpha_\textrm{near}= -1.9(2)$, and is the fit included in Figure~\ref{fig4}. In both analyses, the heating rate frequency scaling is consistent with the prediction from the fluctuation dissipation theorem of $\alpha=-2$ to within the fit uncertainty.

Our results show the feasibility of trapping ions near large dielectric objects with manageable ion heating rates and stray electric fields, opening the door for devices such as trap-integrated high-finesse optical cavities with ion-mirror distances of several hundred microns. Simulations predict that the increase in motional heating due to the dielectric mirror coating in such a cavity should be small~\cite{supplement}. With the potential for high single-ion cooperativity and low motional heating rates, such trap-integrated cavities would be suitable for high-fidelity, high-rate quantum networking applications and cavity quantum electrodynamics studies with trapped ions. \\
 
\begin{acknowledgments}
\textit{Acknowledgments}\textemdash The authors thank Andrew Wilson for important early contributions to the project, Giorgio Zarantonello for design and assembly of the experimental apparatus, and Chin-Wen Chou and Zhimin Liu for maintaining the ultrastable cavity. MB is an associate in the Professional Research Experience Program operated jointly by NIST and the University of Colorado. LS acknowledges support from a National Research Council postdoctoral fellowship. KND acknowledges support from the NSF Quantum Leap Challenge Institute Award OMA - 2016244. This work was supported by the NIST Quantum Network Grand Challenge.

MB and LS performed the experiments, analyzed the data, and wrote the manuscript. LS, with assistance from KND, completed the simulations. DHS designed and fabricated the trap. KND and DHS assembled the fiber on the trap. JS and LS designed and built the apparatus. All authors contributed to experimental design and manuscript editing. DHS and DL secured funding for the work. DL, DHS, and LS supervised the work.
\end{acknowledgments}

\bibliographystyle{apsrev4-2}
\bibliography{refs}{}

@article{berkeland_minimization_1998,
	title = {Minimization of ion micromotion in a {Paul} trap},
	volume = {83},
	copyright = {© 1998 American Institute of Physics.},
	issn = {0021-8979},
	url = {https://aip.scitation.org/doi/abs/10.1063/1.367318},
	doi = {10.1063/1.367318},
	number = {10},
	urldate = {2023-04-13},
	journal = {J. Appl. Phys.},
	author = {Berkeland, D. J. and Miller, J. D. and Bergquist, J. C. and Itano, W. M. and Wineland, D. J.},
	month = apr,
	year = {1998},
	pages = {5025-5033},
}

@article{Wesenberg2008,
  title = {Electrostatics of Surface-Electrode Ion Traps},
  author = {Wesenberg, J. H.},
  year = {2008},
  month = dec,
  journal = {Phys. Rev. A},
  volume = {78},
  number = {6},
  pages = {063410},
  issn = {1050-2947},
  doi = {10.1103/PhysRevA.78.063410},
  urldate = {2012-01-12}
}

@article{Chiaverini2005,
  title = {Surface-{{Electrode Architecture}} for {{Ion-Trap Quantum Information Processing}}},
  author = {Chiaverini, J. and Blakestad, R. B. and Britton, J. and Jost, J. D. and Langer, C. and Leibfried, D. and Ozeri, R. and Wineland, D. J.},
  year = {2005},
  journal = {Quantum Inf. Comput.},
  volume = {5},
  number = {6},
  pages = {419--439},
  doi = {10.26421/QIC5.6-1}
}

@article{home_normalmodes_2011,
    title = {Normal modes of trapped ions in the presence of anharmonic trap potentials},
    author = {Home, J. P. and Hanneke, D. and Jost, J. D. and Leibfried, D. and Wineland, D. J.},
    journal = {New J. Phys.},
    volume = {13},
    pages = {073026},
    year = {2011},
    doi = {10.1088/1367-2630/13/7/073026},
}

@article{moses_race_2023,
  title = {A Race-Track Trapped-Ion Quantum Processor},
  author = {Moses, S. A. and Baldwin, C. H. and others}, 
  journal = {Phys. Rev. X},
  volume = {13},
  issue = {4},
  pages = {041052},
  numpages = {25},
  year = {2023},
  month = {Dec},
  publisher = {American Physical Society},
  doi = {10.1103/PhysRevX.13.041052},
  url = {https://link.aps.org/doi/10.1103/PhysRevX.13.041052}
}

@article{chen_benchmarking_2024,
  title = {Benchmarking a trapped-ion quantum computer with 30 qubits},
  author = {Chen, Jwo-Sy and Nielsen, Erik and Ebert, Matthew and Inlek, Volkan and Wright, Kenneth and Chaplin, Vandiver and Maksymov, Andrii and Paez, Eduardo and Poudel, Amrit and Maunz, Peter and Gamble, John}, 
  journal = {Quantum},
  volume = {8},
  pages = {1516},
  year = {2024},
  month = {Nov},
  doi = {10.22331/q-2024-11-07-1516},
  url = {https://quantum-journal.org/papers/q-2024-11-07-1516/}
}

@article{loschnauer_scalable_2024,
  title = {Scalable, High-Fidelity All-Electronic Control of Trapped-Ion Qubits},
  author = {L\"oschnauer, C.M. and Mosca Toba, J. and Hughes, A.C. and King, S.A. and Weber, M.A. and Srinivas, R. and Matt, R. and Nourshargh, R. and Allcock, D.T.C. and Ballance, C.J. and Matthiesen, C. and Malinowski, M. and Harty, T.P.},
  journal = {PRX Quantum},
  volume = {6},
  issue = {4},
  pages = {040313},
  numpages = {13},
  year = {2025},
  month = {Oct},
  publisher = {American Physical Society},
  doi = {10.1103/h4wk-v31j},
  url = {https://link.aps.org/doi/10.1103/h4wk-v31j}
}

@article{
mehta_inegrated_optics_2016,
	author = {Mehta, Karan K. and Bruzewicz, Colin D. and McConnell, Robert and Ram, Rajeev J. and Sage, Jeremy M. and Chiaverini, John},
	journal = {Nature Nanotech.},
	number = {12},
	pages = {1066--1070},
	title = {Integrated optical addressing of an ion qubit},
	volume = {11},
    url = {https://doi.org/10.1038/nnano.2016.139},
	year = {2016}
}

@article{huntemann_clock_2016,
  title = {Single-Ion Atomic Clock with $3\ifmmode\times\else\texttimes\fi{}{10}^{\ensuremath{-}18}$ Systematic Uncertainty},
  author = {Huntemann, N. and Sanner, C. and Lipphardt, B. and Tamm, Chr. and Peik, E.},
  journal = {Phys. Rev. Lett.},
  volume = {116},
  issue = {6},
  pages = {063001},
  numpages = {5},
  year = {2016},
  month = {Feb},
  publisher = {American Physical Society},
  doi = {10.1103/PhysRevLett.116.063001},
  url = {https://link.aps.org/doi/10.1103/PhysRevLett.116.063001}
}

@article{hausser_clock_2025,
  title = {$^{115}{\mathrm{In}}^{+}\text{\ensuremath{-}}^{172}{\mathrm{Yb}}^{+}$ Coulomb Crystal Clock with $2.5\ifmmode\times\else\texttimes\fi{}{10}^{\ensuremath{-}18}$ Systematic Uncertainty},
  author = {Hausser, H. N. and Keller, J. and Nordmann, T. and Bhatt, N. M. and Kiethe, J. and Liu, H. and Richter, I. M. and von Boehn, M. and Rahm, J. and Weyers, S. and others},
  journal = {Phys. Rev. Lett.},
  volume = {134},
  issue = {2},
  pages = {023201},
  numpages = {6},
  year = {2025},
  month = {Jan},
  publisher = {American Physical Society},
  doi = {10.1103/PhysRevLett.134.023201},
  url = {https://link.aps.org/doi/10.1103/PhysRevLett.134.023201}
}

@Article{niffenegger_multiwav_2020,
author={Niffenegger, R. J.
and Stuart, J.
and Sorace-Agaskar, C.
and Kharas, D.
and Bramhavar, S.
and Bruzewicz, C. D.
and Loh, W.
and Maxson, R. T.
and McConnell, R.
and Reens, D.
and others},
title={Integrated multi-wavelength control of an ion qubit},
journal={Nature},
year={2020},
month={Oct},
day={01},
volume={586},
number={7830},
pages={538-542},
issn={1476-4687},
doi={10.1038/s41586-020-2811-x},
url={https://doi.org/10.1038/s41586-020-2811-x}
}

@Article{mehta_multiion_2020,
author={Mehta, Karan K.
and Zhang, Chi
and Malinowski, Maciej
and Nguyen, Thanh-Long
and Stadler, Martin
and Home, Jonathan P.},
title={Integrated optical multi-ion quantum logic},
journal={Nature},
year={2020},
month={Oct},
day={01},
volume={586},
number={7830},
pages={533-537},
issn={1476-4687},
doi={10.1038/s41586-020-2823-6},
url={https://doi.org/10.1038/s41586-020-2823-6}
}

@phdthesis{Knaack2024,
  title = {Laser-free operations in a mixed-species trapped ion processor},
  author = {Knaack, Hannah M},
  school = {University of Colorado Boulder, 2024},
}

@article{clark_highfidelity_2021,
  title = {High-Fidelity Bell-State Preparation with $^{40}{\mathrm{Ca}}^{+}$ Optical Qubits},
  author = {Clark, Craig R. and Tinkey, Holly N. and Sawyer, Brian C. and Meier, Adam M. and Burkhardt, Karl A. and Seck, Christopher M. and Shappert, Christopher M. and Guise, Nicholas D. and Volin, Curtis E. and Fallek, Spencer D. and others},
  journal = {Phys. Rev. Lett.},
  volume = {127},
  issue = {13},
  pages = {130505},
  numpages = {5},
  year = {2021},
  month = {Sep},
  publisher = {American Physical Society},
  doi = {10.1103/PhysRevLett.127.130505},
  url = {https://link.aps.org/doi/10.1103/PhysRevLett.127.130505}
}

@article{bennett_purification_1996,
  title = {Purification of Noisy Entanglement and Faithful Teleportation via Noisy Channels},
  author = {Bennett, Charles H. and Brassard, Gilles and Popescu, Sandu and Schumacher, Benjamin and Smolin, John A. and Wootters, William K.},
  journal = {Phys. Rev. Lett.},
  volume = {76},
  issue = {5},
  pages = {722--725},
  numpages = {0},
  year = {1996},
  month = {Jan},
  publisher = {American Physical Society},
  doi = {10.1103/PhysRevLett.76.722},
  url = {https://link.aps.org/doi/10.1103/PhysRevLett.76.722}
}

@article{duan_longdistance_2001,
  title = {Long-distance quantum communication with atomic ensembles and linear optics},
  author = {L.-M. Duan and Lukin, M. and Cirac, J and Zoller, P.},
  journal = {Nature},
  volume = {414},
  pages = {413-418},
  year = {2001},
  doi = {10.1038/35106500},
  url = {https://www.nature.com/articles/35106500}
}

@article{duan_scalable_2004,
  author = {L.-M. Duan and Boris B. Blinov and David L. Moehring and Christopher R. Monroe},
  title = {Scalable trapped ion quantum computation with a probabilistic ion-photon mapping},
  journal = {Quantum Inf. Comput.},
  volume = {4},
  number = {3},
  pages = {165-173},
  year = {2004},
  url = {https://doi.org/10.26421/QIC4.3-1},
  doi = {10.26421/QIC4.3-1},
}

@article{kim_tantala_2001,
  author={Jin-Young Kim and Garg, A. and Rymaszewski, E.J. and Toh-Ming Lu},
  journal={IEEE Trans. Compon. Packag. Technol.}, 
  title={High frequency response of amorphous tantalum oxide thin films}, 
  year={2001},
  volume={24},
  number={3},
  pages={526-533},
  doi={10.1109/6144.946502}
}

@Inbook{jain_tantala_2004,
author="Jain, Pushkar
and Rymaszewski, Eugene J.",
title="Applications",
bookTitle="Thin-Film Capacitors for Packaged Electronics",
year="2004",
publisher="Springer US",
address="Boston, MA",
pages="145--153",
isbn="978-1-4419-9144-7",
doi="10.1007/978-1-4419-9144-7_6",
url="https://doi.org/10.1007/978-1-4419-9144-7_6"
}

@article{rempe_measurement_1992,
author = {G. Rempe and R. J. Thompson and H. J. Kimble and R. Lalezari},
journal = {Opt. Lett.},
number = {5},
pages = {363--365},
publisher = {Optica Publishing Group},
title = {Measurement of ultralow losses in an optical interferometer},
volume = {17},
month = {Mar},
year = {1992},
url = {https://opg.optica.org/ol/abstract.cfm?URI=ol-17-5-363},
doi = {10.1364/OL.17.000363},
}

@article{house_analytic_2008,
  title = {Analytic model for electrostatic fields in surface-electrode ion traps},
  author = {House, M. G.},
  journal = {Phys. Rev. A},
  volume = {78},
  issue = {3},
  pages = {033402},
  numpages = {8},
  year = {2008},
  month = {Sep},
  publisher = {American Physical Society},
  doi = {10.1103/PhysRevA.78.033402},
  url = {https://link.aps.org/doi/10.1103/PhysRevA.78.033402}
}

@article{briegel_quantumrepeaters_1998,
  title = {Quantum Repeaters: The Role of Imperfect Local Operations in Quantum Communication},
  author = {Briegel, H.{-}J. and D\"ur, W. and Cirac, J. I. and Zoller, P.},
  journal = {Phys. Rev. Lett.},
  volume = {81},
  issue = {26},
  pages = {5932--5935},
  numpages = {0},
  year = {1998},
  month = {Dec},
  publisher = {American Physical Society},
  doi = {10.1103/PhysRevLett.81.5932},
  url = {https://link.aps.org/doi/10.1103/PhysRevLett.81.5932}
}

@article{wang_singleion_2021,
  title = {Single ion qubit with estimated coherence time exceeding one hour},
  author = {Wang, P. and Luan, C.-Y. and Qiao, M. and Um, M. and Zhang, J. and Wang, Y. and Yuan, X. and Gu, M. and Zhang, J. and Kim, K.},
  journal = {Nat. Commun.},
  volume = {12},
  pages = {233},
  year = {2021},
  doi = {10.1038/s41467-020-20330-w},
  url = {https://www.nature.com/articles/s41467-020-20330-w}
}

@article{Monroe_largescale_2014,
  title = {Large-scale modular quantum-computer architecture with atomic memory and photonic interconnects},
  author = {Monroe, C. and Raussendorf, R. and Ruthven, A. and Brown, K. R. and Maunz, P. and Duan, L.{-}M. and Kim, J.},
  journal = {Phys. Rev. A},
  volume = {89},
  issue = {2},
  pages = {022317},
  numpages = {16},
  year = {2014},
  month = {Feb},
  publisher = {American Physical Society},
  doi = {10.1103/PhysRevA.89.022317},
  url = {https://link.aps.org/doi/10.1103/PhysRevA.89.022317}
}

@article{nichol_elementary_2022,
  title = {An elementary quantum network of entangled optical atomic clocks},
  author = {Nichol, B. C. and Srinivas, R. and Nadlinger, D. P. and Drmota, P. and Main, D. and Araneda, G. and Ballance, C. J. and Lucas, D. M.},
  journal = {Nature},
  volume = {609},
  pages = {689-694},
  year = {2022},
  doi = {10.1038/s41586-022-05088-z},
  url = {https://doi.org/10.1038/s41586-022-05088-z}
}

@article{saha_highfidelity_2024,
	title = {High-fidelity remote entanglement of trapped atoms mediated by time-bin photons},
	author = {Saha, S. and Shalaev, M. and O'Reilly, J. and Goetting, I. and Toh, G. and Kalakuntla, A. and Yu, Y. and Monroe, C.},
	year = {2025},
    journal = {Nat. Commun.},
    volume = {16},
    pages = {2533},
    doi = {s41467-025-57557-4},
    url = {https://www.nature.com/articles/s41467-025-57557-4#Sec14}
}

@article{oreilly_fastphoton_2024,
  title = {Fast Photon-Mediated Entanglement of Continuously Cooled Trapped Ions for Quantum Networking},
  author = {O'Reilly, Jameson and Toh, George and Goetting, Isabella and Saha, Sagnik and Shalaev, Mikhail and Carter, Allison L. and Risinger, Andrew and Kalakuntla, Ashish and Li, Tingguang and Verma, Ashrit and Monroe, Christopher},
  journal = {Phys. Rev. Lett.},
  volume = {133},
  issue = {9},
  pages = {090802},
  numpages = {6},
  year = {2024},
  month = {Aug},
  publisher = {American Physical Society},
  doi = {10.1103/PhysRevLett.133.090802},
  url = {https://link.aps.org/doi/10.1103/PhysRevLett.133.090802}
}

@article{carter_iontrap_2024,
    author = {Carter, Allison L. and O’Reilly, Jameson and Toh, George and Saha, Sagnik and Shalaev, Mikhail and Goetting, Isabella and Monroe, Christopher},
    title = {Ion trap with in-vacuum high numerical aperture imaging for a dual-species modular quantum computer},
    journal = {Rev. Sci. Instrum.},
    volume = {95},
    number = {3},
    pages = {033201},
    year = {2024},
    month = {03},
    doi = {10.1063/5.0180732},
    url = {https://doi.org/10.1063/5.0180732},
}

@article{schupp_interface_2021,
  title = {Interface between Trapped-Ion Qubits and Traveling Photons with Close-to-Optimal Efficiency},
  author = {Schupp, J. and Krcmarsky, V. and Krutyanskiy, V. and Meraner, M. and Northup, T. E. and Lanyon, B. P.},
  journal = {PRX Quantum},
  volume = {2},
  issue = {2},
  pages = {020331},
  numpages = {16},
  year = {2021},
  month = {Jun},
  publisher = {American Physical Society},
  doi = {10.1103/PRXQuantum.2.020331},
  url = {https://link.aps.org/doi/10.1103/PRXQuantum.2.020331}
}

@article{huber_trappedion_2010,
  title = {A trapped-ion local field probe},
  author = {Huber, G. and Ziesel, F. and Poschinger, U. and Singer, K. and Schmidt-Kaler, F.},
  journal = {Appl. Phys. B},
  volume = {100},
  pages = {725-730},
  year = {2010},
  doi = {10.1007/s00340-010-4148-x},
  url = {https://doi.org/10.1007/s00340-010-4148-x}
}

@article{narayanan_electricfield_2011,
    author = {Narayanan, S. and Daniilidis, N. and Möller, S. A. and Clark, R. and Ziesel, F. and Singer, K. and Schmidt-Kaler, F. and Häffner, H.},
    title = {Electric field compensation and sensing with a single ion in a planar trap},
    journal = {J. Appl. Phys.},
    volume = {110},
    number = {11},
    pages = {114909},
    year = {2011},
    month = {12},
    issn = {0021-8979},
    doi = {10.1063/1.3665647},
    url = {https://doi.org/10.1063/1.3665647},
}

@article{takahashi_strongcoupling_2020,
  title = {Strong Coupling of a Single Ion to an Optical Cavity},
  author = {Takahashi, Hiroki and Kassa, Ezra and Christoforou, Costas and Keller, Matthias},
  journal = {Phys. Rev. Lett.},
  volume = {124},
  issue = {1},
  pages = {013602},
  numpages = {5},
  year = {2020},
  month = {Jan},
  publisher = {American Physical Society},
  doi = {10.1103/PhysRevLett.124.013602},
  url = {https://link.aps.org/doi/10.1103/PhysRevLett.124.013602}
}

@article{VanDevender_efficient_2010,
  title = {Efficient Fiber Optic Detection of Trapped Ion Fluorescence},
  author = {VanDevender, A. P. and Colombe, Y. and Amini, J. and Leibfried, D. and Wineland, D. J.},
  journal = {Phys. Rev. Lett.},
  volume = {105},
  issue = {2},
  pages = {023001},
  numpages = {4},
  year = {2010},
  month = {Jul},
  publisher = {American Physical Society},
  doi = {10.1103/PhysRevLett.105.023001},
  url = {https://link.aps.org/doi/10.1103/PhysRevLett.105.023001}
}

@article{Turchette2000,
  title = {Heating of Trapped Ions from the Quantum Ground State},
  author = {Turchette, Q. A. and Kielpinski, David and King, Brian E. and Leibfried, D. and Meekhof, D. M. and Myatt, C. J. and Rowe, M. A. and Sackett, C. A. and Wood, C. S. and Itano, Wayne M. and Monroe, Christopher and Wineland, David J.},
  year = {2000},
  journal = {Phys. Rev. A},
  volume = {61},
  number = {6},
  pages = {063418},
  issn = {1050-2947},
  doi = {10.1103/PhysRevA.61.063418},
}

@article{Sterk2012,
  title = {Photon Collection from a Trapped Ion-Cavity System},
  author = {Sterk, J. D. and Luo, L. and Manning, T. A. and Maunz, P. and Monroe, C.},
  year = {2012},
  month = jun,
  journal = {Phys. Rev. A},
  volume = {85},
  number = {6},
  pages = {062308},
  issn = {1050-2947},
  doi = {10.1103/PhysRevA.85.062308},
  urldate = {2012-06-20},
}

@article{vanrynbach_integrated_2016,
    author = {Van Rynbach, Andre and Maunz, Peter and Kim, Jungsang},
    title = {An integrated mirror and surface ion trap with a tunable trap location},
    journal = {Appl. Phys. Lett.},
    volume = {109},
    number = {22},
    pages = {221108},
    year = {2016},
    month = {12},
    issn = {0003-6951},
    doi = {10.1063/1.4970542},
    url = {https://doi.org/10.1063/1.4970542},
}

@article{Kassa2025,
  title = {How to Integrate a Miniature Optical Cavity in a Linear Ion Trap: {{Shielding}} Dielectrics and Trap Symmetry},
  shorttitle = {How to Integrate a Miniature Optical Cavity in a Linear Ion Trap},
  author = {Kassa, Ezra and Gao, Shaobo and Teh, Soon and {van Dinter}, Dyon and Takahashi, Hiroki},
  year = {2025},
  month = feb,
  journal = {Phys. Rev. Appl.},
  volume = {23},
  number = {2},
  pages = {024038},
  publisher = {American Physical Society},
  doi = {10.1103/PhysRevApplied.23.024038},
}

@article{Knollmann2024,
  title = {Integrated Photonic Structures for Photon-Mediated Entanglement of Trapped Ions},
  author = {Knollmann, F. W. and Clements, E. and Callahan, P. T. and Gehl, M. and Hunker, J. D. and Mahony, T. and McConnell, R. and Swint, R. and {Sorace-Agaskar}, C. and Chuang, I. L. and Chiaverini, J. and Stick, D.},
  year = {2024},
  month = aug,
  journal = {Optica Quantum},
  volume = {2},
  number = {4},
  pages = {230--244},
  publisher = {Optica Publishing Group},
  issn = {2837-6714},
  doi = {10.1364/OPTICAQ.522128},
  urldate = {2024-08-14},
}

@article{Monroe1995a,
  title = {Resolved-{{Sideband Raman Cooling}} of a {{Bound Atom}} to the {{3D Zero-Point Energy}}},
  author = {Monroe, C. and Meekhof, D. M. and King, B. E. and Jefferts, S. R. and Itano, W. M. and Wineland, D. J. and Gould, P.},
  year = {1995},
  month = nov,
  journal = {Phys. Rev. Lett.},
  volume = {75},
  number = {22},
  pages = {4011--4014},
  publisher = {APS},
  issn = {0031-9007},
  doi = {10.1103/PhysRevLett.75.4011},
}

@article{eltony_transparent_2013,
    author = {Eltony, Amira M. and Wang, Shannon X. and Akselrod, Gleb M. and Herskind, Peter F. and Chuang, Isaac L.},
    title = {Transparent ion trap with integrated photodetector},
    journal = {Appl. Phys. Lett.},
    volume = {102},
    number = {5},
    pages = {054106},
    year = {2013},
    month = {02},
    issn = {0003-6951},
    doi = {10.1063/1.4790843},
    url = {https://doi.org/10.1063/1.4790843},
}

@article{ong_probingsurface_2020,
  title = {Probing surface charge densities on optical fibers with a trapped ion},
  author = {Ong, F. R. and Sch\"{u}ppert, K. and Jobez, P. and Teller, M. and Ames, B. and Fioretto, D. A. and Friebe, K. and Lee, M. and Colombe, Y. and Blatt, R. and Northup, T. E.},
  journal = {New J. Phys.},
  volume = {22},
  pages = {063018},
  year = {2020},
  doi = {10.1088/1367-2630/ab8af9},
}

@article{leibfried_quantum_2003,
	title = {Quantum dynamics of single trapped ions},
	volume = {75},
	issn = {0034-6861, 1539-0756},
	url = {https://link.aps.org/doi/10.1103/RevModPhys.75.281},
	doi = {10.1103/RevModPhys.75.281},
	language = {en},
	number = {1},
	urldate = {2022-05-02},
	journal = {Rev. Mod. Phys.},
	author = {Leibfried, D. and Blatt, R. and Monroe, C. and Wineland, D.},
	month = mar,
	year = {2003},
	pages = {281--324},
}

@article{teller_heating_2021,
  title = {Heating of a Trapped Ion Induced by Dielectric Materials},
  author = {Teller, Markus and Fioretto, Dario A. and Holz, Philip C. and Schindler, Philipp and Messerer, Viktor and Sch\"uppert, Klemens and Zou, Yueyang and Blatt, Rainer and Chiaverini, John and Sage, Jeremy and Northup, Tracy E.},
  journal = {Phys. Rev. Lett.},
  volume = {126},
  issue = {23},
  pages = {230505},
  numpages = {6},
  year = {2021},
  month = {Jun},
  publisher = {American Physical Society},
  doi = {10.1103/PhysRevLett.126.230505},
  url = {https://link.aps.org/doi/10.1103/PhysRevLett.126.230505}
}

@article{sorensen_quantum_1999,
  title = {Quantum Computation with Ions in Thermal Motion},
  author = {S\o{}rensen, Anders and M\o{}lmer, Klaus},
  journal = {Phys. Rev. Lett.},
  volume = {82},
  issue = {9},
  pages = {1971--1974},
  numpages = {0},
  year = {1999},
  month = {Mar},
  publisher = {American Physical Society},
  doi = {10.1103/PhysRevLett.82.1971},
  url = {https://link.aps.org/doi/10.1103/PhysRevLett.82.1971}
}

@article{sorensen_entanglement_2000,
  title = {Entanglement and quantum computation with ions in thermal motion},
  author = {S\o{}rensen, Anders and M\o{}lmer, Klaus},
  journal = {Phys. Rev. A},
  volume = {62},
  issue = {2},
  pages = {022311},
  numpages = {11},
  year = {2000},
  month = {Jul},
  publisher = {American Physical Society},
  doi = {10.1103/PhysRevA.62.022311},
  url = {https://link.aps.org/doi/10.1103/PhysRevA.62.022311}
}

@article{brownnutt_iontrap_2015,
  title = {Ion-trap measurements of electric-field noise near surfaces},
  author = {Brownnutt, M. and Kumph, M. and Rabl, P. and Blatt, R.},
  journal = {Rev. Mod. Phys.},
  volume = {87},
  issue = {4},
  pages = {1419--1482},
  numpages = {64},
  year = {2015},
  month = {Dec},
  publisher = {American Physical Society},
  doi = {10.1103/RevModPhys.87.1419},
  url = {https://link.aps.org/doi/10.1103/RevModPhys.87.1419}
}

@article{kumph_electric-field_2016,
	title = {Electric-field noise above a thin dielectric layer on metal electrodes},
	volume = {18},
	url = {https://dx.doi.org/10.1088/1367-2630/18/2/023020},
	doi = {10.1088/1367-2630/18/2/023020},
	number = {2},
	journal = {New J. Phys.},
	author = {Kumph, Muir and Henkel, Carsten and Rabl, Peter and Brownnutt, Michael and Blatt, Rainer},
	month = feb,
	year = {2016},
	pages = {023020},
}

@misc{supplement,
    author = {},
    note = {See Supplemental Material below for specifics on the optical fiber attachment, radial stray field drift, details of the stray charge density model, and additional simulations of the dielectric-induced heating rate. The Supplemental Material includes additional references [76-84].}
}

@misc{nist_disclaimer,
    author = {},
    note = {Certain equipment, instruments, software, or materials are identified in this paper in order to specify the experimental procedure adequately. Such identification is not intended to imply recommendation or endorsement of any product or service by NIST, nor is it intended to imply that the materials or equipment identified are necessarily the best available for the purpose.}
}

@article{todaro_statereadout_2021,
  title = {State Readout of a Trapped Ion Qubit Using a Trap-Integrated Superconducting Photon Detector},
  author = {Todaro, S. L. and Verma, V. B. and McCormick, K. C. and Allcock, D. T. C. and Mirin, R. P. and Wineland, D. J. and Nam, S. W. and Wilson, A. C. and Leibfried, D. and Slichter, D. H.},
  journal = {Phys. Rev. Lett.},
  volume = {126},
  issue = {1},
  pages = {010501},
  numpages = {7},
  year = {2021},
  month = {Jan},
  publisher = {American Physical Society},
  doi = {10.1103/PhysRevLett.126.010501},
  url = {https://link.aps.org/doi/10.1103/PhysRevLett.126.010501}
}

@article{kim_surfaceelectrode_2011,
    author = {Kim, Tony Hyun and Herskind, Peter F. and Chuang, Isaac L.},
    title = {Surface-electrode ion trap with integrated light source},
    journal = {Appl. Phys. Lett.},
    volume = {98},
    number = {21},
    pages = {214103},
    year = {2011},
    month = {05},
    issn = {0003-6951},
    doi = {10.1063/1.3593496},
    url = {https://doi.org/10.1063/1.3593496},
}

@article{takahashi_integrated_2013,
    author = {Takahashi, Hiroki and Wilson, Alex and Riley-Watson, Andrew and Oručević, Fedja and Seymour-Smith, Nicolas and Keller, Matthias and Lange, Wolfgang},
    title = {An integrated fiber trap for single-ion photonics},
    journal = {New. J. Phys.},
    volume = {15},
    pages = {053011},
    year = {2013},
    doi = {10.1088/1367-2630/15/5/053011},
}

@article{jechow_quantum_2011,
author = {A. Jechow and E. W. Streed and B. G. Norton and M. J. Petrasiunas and D. Kielpinski},
journal = {Opt. Lett.},
number = {8},
pages = {1371--1373},
publisher = {Optica Publishing Group},
title = {Wavelength-scale imaging of trapped ions using a phase Fresnel lens},
volume = {36},
month = {Apr},
year = {2011},
url = {https://opg.optica.org/ol/abstract.cfm?URI=ol-36-8-1371},
doi = {10.1364/OL.36.001371},
}

@article{gulde_simple_2001,
    author = {Gulde, S. and Rotter, D. and Barton, P. and Schmidt-Kaler, F. and Blatt, R. and Hogervorst, W.},
    title = {Simple and efficient photo-ionization loading of ions for precision ion-trapping experiments},
    journal = {Appl. Phys. B},
    volume = {73},
    pages = {861-863},
    year = {2001},
    month = {03},
    doi = {10.1007/s003400100749},
    url = {https://link.springer.com/article/10.1007/s003400100749},
}

@Inbook{jain_thinfilm_2004,
author="Jain, Pushkar
and Rymaszewski, Eugene J.",
title="Electrical Characterization",
bookTitle="Thin-Film Capacitors for Packaged Electronics",
year="2004",
publisher="Springer US",
address="Boston, MA",
pages="91--119",
isbn="978-1-4419-9144-7",
doi="10.1007/978-1-4419-9144-7_4",
url="https://doi.org/10.1007/978-1-4419-9144-7_4"
}

@article{marshall_highstability_2025,
  title = {High-Stability Single-Ion Clock with $5.5\ifmmode\times\else\texttimes\fi{}{10}^{\ensuremath{-}19}$ Systematic Uncertainty},
  author = {Marshall, Mason C. and Castillo, Daniel A. Rodriguez and Arthur-Dworschack, Willa J. and Aeppli, Alexander and Kim, Kyungtae and Lee, Dahyeon and Warfield, William and Hinrichs, Joost and Nardelli, Nicholas V. and Fortier, Tara M. and others},
  journal = {Phys. Rev. Lett.},
  volume = {135},
  issue = {3},
  pages = {033201},
  numpages = {6},
  year = {2025},
  month = {Jul},
  publisher = {American Physical Society},
  doi = {10.1103/hb3c-dk28},
  url = {https://link.aps.org/doi/10.1103/hb3c-dk28}
}

@article{smith_singlequbit_2025,
  title = {Single-Qubit Gates with Errors at the ${10}^{\ensuremath{-}7}$ Level},
  author = {Smith, M. C. and Leu, A. D. and Miyanishi, K. and Gely, M. F. and Lucas, D. M.},
  journal = {Phys. Rev. Lett.},
  volume = {134},
  issue = {23},
  pages = {230601},
  numpages = {8},
  year = {2025},
  month = {Jun},
  publisher = {American Physical Society},
  doi = {10.1103/42w2-6ccy},
  url = {https://link.aps.org/doi/10.1103/42w2-6ccy}
}

@article{Milburn2000,
  title = {Ion Trap Quantum Computing with Warm Ions},
  author = {Milburn, G. J. and Schneider, S. and James, D. F V},
  year = {2000},
  journal = {Fortschr. Phys.},
  volume = {48},
  pages = {801--810},
  doi = {10.1002/1521-3978(200009)48:9/11<801::AID-PROP801>3.0.CO;2-1},
}

@article{Pfeifer2022,
  title = {Achievements and Perspectives of Optical Fiber {{Fabry}}--{{Perot}} Cavities},
  author = {Pfeifer, H. and Ratschbacher, L. and Gallego, J. and Saavedra, C. and Fa{\ss}bender, A. and {von Haaren}, A. and Alt, W. and Hofferberth, S. and K{\"o}hl, M. and Linden, S. and Meschede, D.},
  year = {2022},
  month = jan,
  journal = {Appl. Phys. B},
  volume = {128},
  number = {2},
  pages = {29},
  issn = {1432-0649},
  doi = {10.1007/s00340-022-07752-8},
  urldate = {2023-08-21},
  langid = {english}
}

@article{Bernien2017,
  title = {Probing Many-Body Dynamics on a 51-Atom Quantum Simulator},
  author = {Bernien, Hannes and Schwartz, Sylvain and Keesling, Alexander and Levine, Harry and Omran, Ahmed and Pichler, Hannes and Choi, Soonwon and Zibrov, Alexander S. and Endres, Manuel and Greiner, Markus and Vuleti{\'c}, Vladan and Lukin, Mikhail D.},
  year = {2017},
  month = nov,
  journal = {Nature},
  volume = {551},
  number = {7682},
  pages = {579--584},
  publisher = {Nature Publishing Group},
  issn = {0028-0836},
  doi = {10.1038/nature24622}
}

@article{Day2021,
  title = {A Micro-Optical Module for Multi-Wavelength Addressing of Trapped Ions},
  author = {Day, Matthew L and Choonee, Kaushal and Chaboyer, Zachary and Gross, Simon and Withford, Michael J and Sinclair, Alastair G and Marshall, Graham D},
  year = {2021},
  month = feb,
  journal = {Quantum Sci. Technol.},
  volume = {6},
  number = {2},
  pages = {024007},
  publisher = {IOP Publishing},
  issn = {2058-9565},
  doi = {10.1088/2058-9565/abdf38}
}

@techreport{Bado2015,
  title = {Fused {{Silica Ion Trap Chip}} with {{Efficient Optical Collection System}} for {{Timekeeping}}, {{Sensing}}, and {{Emulation}}},
  author = {Bado, Philippe},
  year = {2015},
  number = {FA9550-12-C-0074},
}

@article{Pagano2018,
  title = {Cryogenic Trapped-Ion System for Large Scale Quantum Simulation},
  author = {Pagano, G and Hess, P W and Kaplan, H B and Tan, W L and Richerme, P and Becker, P and Kyprianidis, A and Zhang, J and Birckelbaw, E and Hernandez, M R and Wu, Y and Monroe, C},
  year = {2018},
  month = oct,
  journal = {Quantum Sci. Technol.},
  volume = {4},
  number = {1},
  pages = {014004},
  publisher = {IOP Publishing},
  issn = {2058-9565},
  doi = {10.1088/2058-9565/aae0fe}
}

@article{kay_carbon_2024,
author = {D. J. Kay and S. J. Snowden and G. Stutter and M. K. Keller},
journal = {Opt. Express},
number = {24},
pages = {43654--43662},
publisher = {Optica Publishing Group},
title = {Micro-mirror laser machining for ultra-low birefringence cavities},
volume = {32},
month = {Nov},
year = {2024},
url = {https://opg.optica.org/oe/abstract.cfm?URI=oe-32-24-43654},
doi = {10.1364/OE.542119},
}

@article{gao_profile_2025,
author = {Shaobo Gao and Vishnu Kavungal and Shuma Oya and Daichi Okuno and Ezra Kassa and William J. Hughes and Peter Horak and Hiroki Takahashi},
journal = {Opt. Express},
number = {18},
pages = {39009--39022},
publisher = {Optica Publishing Group},
title = {Profile control of fiber-based micro-mirrors using adaptive laser shooting with in situ imaging},
volume = {33},
month = {Sep},
year = {2025},
url = {https://opg.optica.org/oe/abstract.cfm?URI=oe-33-18-39009},
doi = {10.1364/OE.564341}
}

@article{teller_integrating_2023,
    author = {Teller, Markus and Messerer, Viktor and Schüppert, Klemens and Zou, Yueyang and Fioretto, Dario A. and Galli, Maria and Holz, Philip C. and Reichel, Jakob and Northup, Tracy E.},
    title = {Integrating a fiber cavity into a wheel trap for strong ion–cavity coupling},
    journal = {AVS Quantum Sci.},
    volume = {5},
    number = {1},
    pages = {012001},
    year = {2023},
    month = {01},
    issn = {2639-0213},
    doi = {10.1116/5.0121534},
    url = {https://doi.org/10.1116/5.0121534},
}

@article{hunger_fiber_2010,
author = {Hunger, D. and Steinmetz, T. and Colombe, Y. and Deutsch, C. and H\"{a}nsch, T.W. and Reichel, J.},
title = {A fiber Fabry-Perot cavity with high finesse},
journal = {New J. Phys.},
volume = {12},
pages = {065038},
year = {2010},
doi = {10.1088/1367-2630/12/6/065038},
}

@article{gao_optimization_2023,
  title = {Optimization of Scalable Ion-Cavity Interfaces for Quantum Photonic Networks},
  author = {Gao, Shaobo and Blackmore, Jacob A. and Hughes, William J. and Doherty, Thomas H. and Goodwin, Joseph F.},
  journal = {Phys. Rev. Appl.},
  volume = {19},
  issue = {1},
  pages = {014033},
  numpages = {17},
  year = {2023},
  month = {Jan},
  publisher = {American Physical Society},
  doi = {10.1103/PhysRevApplied.19.014033},
  url = {https://link.aps.org/doi/10.1103/PhysRevApplied.19.014033}
}

@article{kobel_deterministic_2021,
title = {Deterministic spin-photon entanglement from a trapped ion in a fiber Fabry-Perot cavity},
author = {Kobel, Pascal and Breyer, Moritz and K\"{o}hl, Michael},
journal = {npj Quantum Inf.},
volume = {7},
pages={6},
year= {2021},
url = {https://www.nature.com/articles/s41534-020-00338-2}
}

@inproceedings{quantinuum_integrated_2023,
author = {Rezlind Bushati and others},
title = {{Trapped-ion quantum computing with integrated photonics}},
volume = {12794},
booktitle = {Emerging Applications in Silicon Photonics IV},
editor = {Callum G. Littlejohns and Marc Sorel},
organization = {International Society for Optics and Photonics},
publisher = {SPIE},
pages = {127940G},
year = {2023},
doi = {10.1117/12.3008419},
URL = {https://doi.org/10.1117/12.3008419}
}

@unpublished{apolin_recoilinduced_2025,
      title={Recoil-induced errors and their correction in photon-mediated entanglement between atom qubits}, 
      author={Jan Apolín and David P. Nadlinger},
      year={2025},
      eprint={2503.16837},
      archivePrefix={arXiv},
      primaryClass={quant-ph},
      url={https://arxiv.org/abs/2503.16837}, 
}

@article{kikura_recoil_2025,
  title = {Taming the Recoil Effect in Cavity-Assisted Quantum Interconnects},
  author = {Kikura, Seigo and Inoue, Ryotaro and Yamasaki, Hayata and Goban, Akihisa and Sunami, Shinichi},
  journal = {PRX Quantum},
  volume = {6},
  issue = {4},
  pages = {040351},
  numpages = {16},
  year = {2025},
  month = {Dec},
  publisher = {American Physical Society},
  doi = {10.1103/njh8-q7gb},
  url = {https://link.aps.org/doi/10.1103/njh8-q7gb}
}

@phdthesis{susanna_scalable_2020,
    author = {Todaro, Susanna},
    title = {Scalable State Detection and Fast Transport of Trapped-Ion Qubits for Quantum Computing},
    school = {University of Colorado at Boulder},
    year = {2020},
}

@phdthesis{roos_controlling_2000,
    author = {Roos, Christian},
    title = {Controlling the Quantum State of Trapped Ions},
    school = {University of Innsbruck},
    year = {2000},
}

@article{dubielzig_ultralowvibration_2021,
    author = {Dubielzig, T. and Halama, S. and Hahn, H. and Zarantonello, G. and Niemann, M. and Bautista-Salvador, A. and Ospelkaus, C.},
    title = {Ultra-low-vibration closed-cycle cryogenic surface-electrode ion trap apparatus},
    journal = {Rev. Sci. Instrum.},
    volume = {92},
    number = {4},
    pages = {043201},
    year = {2021},
    month = {04},
    issn = {0034-6748},
    doi = {10.1063/5.0024423},
    url = {https://doi.org/10.1063/5.0024423},
}

@article{garrett_measuring_2020,
  title = {Measuring the effect of electrostatic patch potentials in Casimir force experiments},
  author = {Garrett, Joseph L. and Kim, Jongbum and Munday, Jeremy N.},
  journal = {Phys. Rev. Research},
  volume = {2},
  issue = {2},
  pages = {023355},
  numpages = {5},
  year = {2020},
  month = {Jun},
  publisher = {American Physical Society},
  doi = {10.1103/PhysRevResearch.2.023355},
  url = {https://link.aps.org/doi/10.1103/PhysRevResearch.2.023355}
}

@article{gu_hybrid_2024,
    author={Fenglei Gu and Shankar G Menon and David Maier and Antariksha Das and Tanmoy Chakraborty and Wolfgang Tittel and Hannes Bernien and Johannes Borregaard},
    year={2025}, 
    journal = {npj Quantum Inf.},
    volume = {11},
    issue = {1},
    pages = {182},
    doi = {10.1038/s41534-025-01119-5},
    url = {https://doi.org/10.1038/s41534-025-01119-5}
}

@article{brekenfeld_quantum_2020,
      title={A quantum network node with crossed optical fibre cavities}, 
      author={Brekenfeld, Manuel and Niemietz, Dominik and Christesen, Joseph Dale and Rempe, Gerhard},
      year={2020},
      journal = {Nature},
      volume = {16},
      pages = {647-651},
      url = {https://www.nature.com/articles/s41567-020-0855-3},
}

@article{brady_integration_2011,
      title={Integration of fluorescence collection optics with a microfabricated surface electrode ion trap}, 
      author={Brady, G.R. and Ellis, A. R. and Moehring, D. L. and Stick, D. and Highstrete, C. and Fortier, K. M. and Blain, M. G. and Haltli, R. A. and Cruz-Cabrera, A. A. and Briggs, R. D. and Wendt, J. R. and Carter, T. R. and Samora, S. and Kenne, S. A.},
      year={2011},
      journal = {Appl. Phys. B},
      volume = {103},
      pages = {801-808},
      url = {https://doi.org/10.1007/s00340-011-4453-z},
}

@article{deshmukh_dielectric_2023,
author = {Chetan Deshmukh and Eduardo Beattie and Bernardo Casabone and Samuele Grandi and Diana Serrano and Alban Ferrier and Philippe Goldner and David Hunger and Hugues de Riedmatten},
journal = {Optica},
number = {10},
pages = {1339--1344},
publisher = {Optica Publishing Group},
title = {Detection of single ions in a nanoparticle coupled to\&\#x00A0;a fiber cavity},
volume = {10},
month = {Oct},
year = {2023},
url = {https://opg.optica.org/optica/abstract.cfm?URI=optica-10-10-1339},
doi = {10.1364/OPTICA.491692},
}

@article{grinkemeyer_errordetected_2025,
author = {Grinkemeyer, Brandon and Guardado-Sanchez, Elmer and Dimitrova, Ivana and Shchepanovich, Danilo and Mandopoulou, G, Eirini and Borregaard, Johannes and Vuleti\'{c}, Vladan and Lukin, Mikhail D.},
journal = {Science},
title = {Error-detected quantum operations with neutral atoms mediated by an optical cavity},
number = {6740},
pages = {1301-1305},
volume = {387},
year = {2025},
url = {https://www.science.org/doi/full/10.1126/science.adr7075},
doi = {10.1126/science.adr7075},
}

@article{riedel_deterministic_2017,
  title = {Deterministic Enhancement of Coherent Photon Generation from a Nitrogen-Vacancy Center in Ultrapure Diamond},
  author = {Riedel, Daniel and S\"ollner, Immo and Shields, Brendan J. and Starosielec, Sebastian and Appel, Patrick and Neu, Elke and Maletinsky, Patrick and Warburton, Richard J.},
  journal = {Phys. Rev. X},
  volume = {7},
  issue = {3},
  pages = {031040},
  numpages = {8},
  year = {2017},
  month = {Sep},
  publisher = {American Physical Society},
  doi = {10.1103/PhysRevX.7.031040},
  url = {https://link.aps.org/doi/10.1103/PhysRevX.7.031040}
}

@article{deslauriers_scaling_2006,
      title={Scaling and Suppression of Anomalous Heating in Ion Traps}, 
      author={Deslauriers, L. and Olmschenk, S. and Stick, D. and Hensinger, W. K. and Sterk, J. and Monroe, C.},
      year={2006},
      journal = {Phys. Rev. Lett.},
      volume = {97},
      url = {https://doi.org/10.1103/PhysRevLett.97.103007},
}

@article{chen_design_2025,
doi = {10.1088/2058-9565/ae379e},
url = {https://doi.org/10.1088/2058-9565/ae379e},
year = {2026},
month = {jan},
publisher = {IOP Publishing},
volume = {11},
number = {1},
pages = {015045},
author = {Chen, Wei-Bin and Fang, Ding and Zhang, Cheng-Hao and Cui, Jin-Ming and Huang, Yun-Feng and Li, Chuan-Feng and Guo, Guang-Can},
title = {Design and fabrication of metal-shielded fiber-cavity mirrors for ion-trap systems},
journal = {Quantum Sci. Technol.},
}

@article{fang_cavity_2025,
  title = {Cavity-assisted thermometry for a single trapped ion},
  author = {Fang, Ding and Chen, Wei-Bin and Zhang, Cheng-Hao and Cui, Jin-Ming and Huang, Yun-Feng and Li, Chuan-Feng and Guo, Guang-Can},
  journal = {Phys. Rev. Appl.},
  volume = {24},
  issue = {2},
  pages = {024037},
  numpages = {8},
  year = {2025},
  month = {Aug},
  publisher = {American Physical Society},
  doi = {10.1103/vjl6-crbg},
  url = {https://link.aps.org/doi/10.1103/vjl6-crbg}
}

\clearpage
\onecolumngrid

\begin{center}
\textbf{\large Supplemental Material for ``Compatibility of Trapped Ions and Dielectrics at Cryogenic Temperatures''}

\vspace{0.5em}

\normalsize
M.~Bruff$^{1,2,*}$, L.~Sonderhouse$^{1,*}$, K.~N.~David$^{1,2}$, J.~Stuart$^{1,\dagger}$, D.~H.~Slichter$^{1}$, and D.~Leibfried$^{1}$

\vspace{0.5em}

\small
\textit{$^{1}$National Institute of Standards and Technology, 325 Broadway, Boulder, CO 80305, USA} \\
\textit{$^{2}$Department of Physics, University of Colorado, Boulder, CO 80309, USA}

\vspace{1em}

\end{center}

\footnotetext{These authors contributed equally to this work.}
\footnotetext{Current address: IonQ, 4505 Campus Drive, College Park, MD \newline \noindent20740}

\twocolumngrid

\setcounter{figure}{0}
\setcounter{equation}{0}

\subsection{Optical fiber attachment}
An $\approx$2\,cm long, $125\,\mu$m diameter 780HP optical fiber is mounted on the trap, as shown in Figure \ref{fig:fiber}. To prepare the fiber, we heat strip the protective buffer and remove any acrylate residue with isopropanol. We cleave the bare fiber, resulting in a fiber facet angled at approximately $1^\circ$. The fiber is secured directly on the trap using a sleeve formed by 25 $\mu$m diameter gold wire bonds that span adjacent grounded electrodes. This stabilizes the fiber's radial position ($y$ and $z$) relative to the the trap.

The trap is mounted with 97\,\% In, 3\,\% Ag solder onto a Au-plated copper puck that also holds the surrounding printed circuit board (PCB). The solder was melted using Nanofoil to attach the trap~\cite{nist_disclaimer, Knaack2024}. The PCB contains resistors and capacitors for filtering signals that drive the dc electrodes. Placing these components close to the trap reduces electrical pickup on the trap electrodes, while keeping them at cryogenic temperatures minimizes Johnson noise that could lead to ion motional heating. The fiber extends $\approx 15$\,mm past the edge of the trap chip, with the other end resting on a grounded copper pedestal that rises from the PCB to the height of the trap. This helps keep the fiber horizontal on the trap chip. The $x$ position of the fiber is set under a high resolution microscope, aligned to have the facet nominally at the end of the RF electrodes. Once aligned, the $x$ position of the fiber is fixed by  securing the far end of the fiber to the copper pedestal with a UV-cured optical adhesive.

\renewcommand{\figurename}{FIG.}
\renewcommand{\thefigure}{S\arabic{figure}}
\begin{figure}
\includegraphics[width=8.4cm]{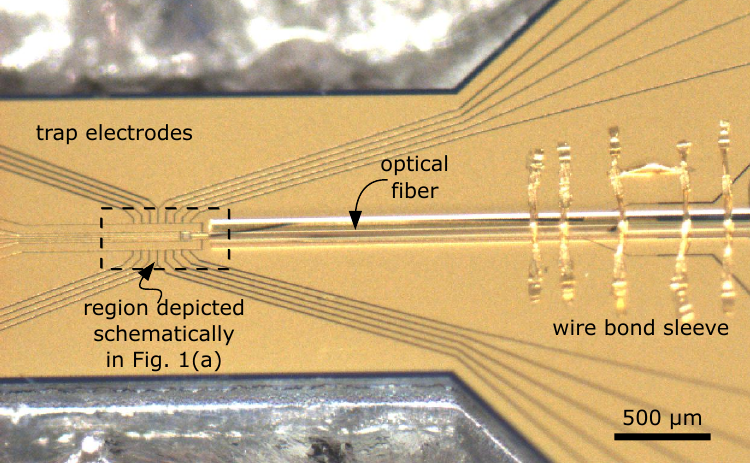}
\caption{Photograph of the surface electrode ion trap with an optical fiber attached approximately parallel to and centered on the trap axis. The segment is cleaved to have a clean facet on the side near the trapping region. It is kept in place with a sleeve of wire bonds tacked to the adjacent grounded electrodes and further constrained off-chip with a small amount of adhesive $\approx 2$\,cm away from the trapping region (not shown).}
\label{fig:fiber}
\end{figure}

\subsection{Fiber position calibration\label{sec:fiberpos}}
We determine the distance from the fiber to the ion $d$ using transmission and scattering of the 729 nm laser light that drives the quadrupole transition, which has a $\approx\,30\,\mu$m beam waist at the ion. The position of the ion is determined by maximizing the Rabi frequency of the $\ket{\downarrow}\leftrightarrow\ket{\uparrow}$ transition using this laser light. The position of the optical fiber is then found relative to the measured ion position. A $z$ scan of the laser beam enables identification of the fiber's top edge via a knife-edge-type measurement. Given the known fiber diameter and ion height above the trap~\cite{Chiaverini2005, Wesenberg2008, house_analytic_2008}, the relative height of the fiber edge determines a gap between the trap and bottom of the fiber of $4(4)\,\mu$m. This value also sets the height of the fiber center, which is used for the following position calibration in the $x$-$y$ plane.

To determine the $y$-offset of the fiber relative to the $y$ position of the ion, the 729 nm laser beam is sent in a different optical port of the chamber, to be incident directly toward the fiber facet, and scanned in $y$. The edges of the fiber face are fit as maxima in light scatter as measured in an appropriate region of interest on the camera, and are determined relative to the $y$ position of the ion. The measured offset is found to be $y_\textrm{fib}=-20(3)\,\mu$m. Due to limited optical access, scanning the laser beam horizontally along the $x$ axis for a knife-edge measurement of the fiber's front face is not possible. Instead, we find one edge of the front face from the transmitted power with the laser light propagating as shown in Figure 1 of the main text. The opposite edge is found after swapping the laser beam position so that it propagates along the direction labeled ``cooling/detection'' in Figure 1 of the main text. In both cases, the edge is measured relative to the $x$ position of an ion trapped at a known position with the axial stray field compensated. The two diagonal knife-edge measurements together constrain the $x$ position of the fiber to be $1(4)\,\mu$m to the right of the rf electrode edge (see Figure 1a), and a tilt of the fiber in the $x$-$y$ plane relative to the $x$ axis of $-1(3)\,^{\circ}$. We use this $x$ position relative to the trap coordinate system to provide the ion-fiber distances presented in the main text. The COMSOL model uses these $x$ and $y$ positions for the fiber face, but does not include a tilt angle since it is consistent with zero.

\subsection{Compensation fields}
The compensation voltages are calculated based on electrostatic modeling of the trap electrode geometry to produce a known, spatially uniform electric field at the ion position without changing the potential curvature. The compensation voltages required to compensate in the three directions yield the stray electric field. For the radial directions, the ion is moved to the rf null line (minimizing rf micromotion) by iteratively reducing either the motional excitation that occurs from modulating the pseudopotential amplitude at the secular frequency~\cite{home_normalmodes_2011} or the micromotion sideband Rabi frequency of the $\ket{\downarrow}\leftrightarrow\ket{\uparrow}$ transition~\cite{berkeland_minimization_1998}. $E_x$ is found by relaxing the axial confinement and measuring the resulting ion displacement on the camera. For small displacements $\Delta x$ and assuming $E_x$ is uniform, $E_x=m\omega_x^2\Delta x/q$, where $\omega_x$ is the axial mode frequency. Alternatively, an $x$ compensation voltage is added to the dc electrodes until the ion displacement is minimized. We have confirmed that $E_x$ extracted from these methods agree for small ion displacements of a few $\mu$m.

\subsection{Stray charge density modeling}

The stray field can be modeled by a surface charge distribution on the optical fiber using finite element analysis. The model uses the surface electrode trap geometry with all electrodes grounded. The optical fiber is modeled as a 200\,$\mu$m long, 125$\,\mu$m diameter cylinder with the assigned material properties of SiO$_2$, and is placed relative to the trap electrodes using the measured optical fiber position described in the previous section. We model the stray electric field using five fit parameters. Two parameters model the surface charge distribution on the optical fiber: a homogeneous surface charge density over the entire optical fiber, $A$, and an additional surface charge difference between the +$\hat{y}$ and -$\hat{y}$ cylindrical side walls of the optical fiber, $D$. The remaining three free parameters are in the form of a uniform offset of the stray field, $\vec{E}_0$, which account for an unknown charge distribution on the surface trap to first order. The full fit is given by: 
\makeatletter 
\renewcommand{\theequation}{S\@arabic\c@equation}
\makeatother
\begin{align}
\vec{E}(\vec{x}) = \vec{E}_0 &+ A \, \vec{E}_{\mathrm{front}}(\vec{x}) + \left(A+\frac{D}{2}\right) \,\vec{E}_{+\hat{y}}(\vec{x}) \nonumber \\
&+ \left(A-\frac{D}{2}\right) \,\vec{E}_{-\hat{y}}(\vec{x})\,,
\end{align}
where $\vec{E}_{\mathrm{front}}$, $\vec{E}_{+\hat{y}}$, and $\vec{E}_{-\hat{y}}$ are the electric fields simulated for a fixed surface charge density on the front, $+\hat{y}$ side wall, and $-\hat{y}$ side wall of the optical fiber, respectively. Since the electric field scales linearly with the surface charge density at the ion, we can assign a surface charge density $\sigma$ from $A$ and $D$.
This model uses the fewest free parameters while still adequately describing the observed stray field. The offset field reduces the fit residuals, but introduces correlations between the offsets and the surface charge density on the optical fiber. For the fit we present here, the extracted surface charge parameters change by less than 20\,\% when setting $\vec{E}_0=0$.  

\subsection{Radial stray field drift}
\begin{figure}
\includegraphics[width=8.4cm]{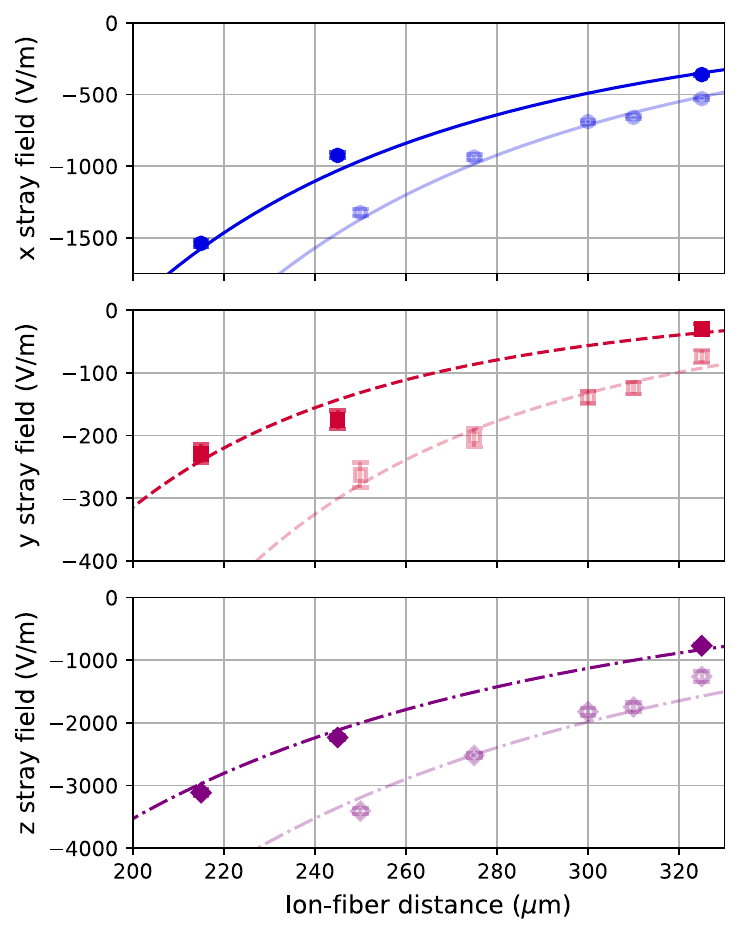}
\caption{Stray electric field in the $x$ ($E_x$, blue circles), $y$ ($E_y$, red squares), and $z$ ($E_z$, purple diamonds) directions from data run VI in March 2025 (dark, solid). Field components are found within the same day at each ion-fiber distance. The fitted $x$, $y$, and $z$ field components are shown as lines in solid blue, dashed red, and dot-dashed purple, respectively. These data and fit are compared to those presented in the main text from earlier data run II in Oct. 2024 (faded), showing an overall decrease in field strength and relaxation of stray charge over time. See text for comparison of fit parameters. All error bars represent 68\,\% confidence intervals.}
\label{fig:3d_drifts}
\end{figure}

In addition to the $x$ stray field drift shown in Figure 2(b), we characterize the drift of all three components of the stray field by comparing those from November 2024 (data run II in Figure 2(a)) to a subsequent data run (hereafter data run VI) from March 2025. The data, shown in Figure ~\ref{fig:3d_drifts}, show a gradual decrease of the stray field in all three directions. The average fractional decrease in $|\vec{E}|$ is approximately 10\,\% per month.

The data in run VI can be used to fit the surface charge density on the optical fiber, using the same model as discussed in the previous section. We find that for data run VI, $\mathrm{\sigma_{f} = 6.9(4)\,e/\mu m^2}$, $\mathrm{\sigma_{+\hat{y}} = 28(5)\,e/\mu m^2}$, and $\mathrm{\sigma_{-\hat{y}} = -14(5) \,e/\mu m^2}$, where $e$ is the elementary charge, while $\vec{E}_0 = (160 \pm 40, 20 \pm 50, 770 \pm 140)$\,V/m. This is compared with data run II, where $\mathrm{\sigma_{f} = 9.5(6)\,e/\mu m^2}$, $\mathrm{\sigma_{+\hat{y}} = 47(9)\,e/\mu m^2}$, and $\mathrm{\sigma_{-\hat{y}} = -28(9) \,e/\mu m^2}$, while $\vec{E}_0 = (190 \pm 60, 20 \pm 50, 620 \pm 210)$\,V/m. This is consistent with an overall slow dissipation of surface charge over time.

\subsection{Further simulations of dielectric-induced heating rates}

\begin{figure}
\includegraphics[width=8.5cm]{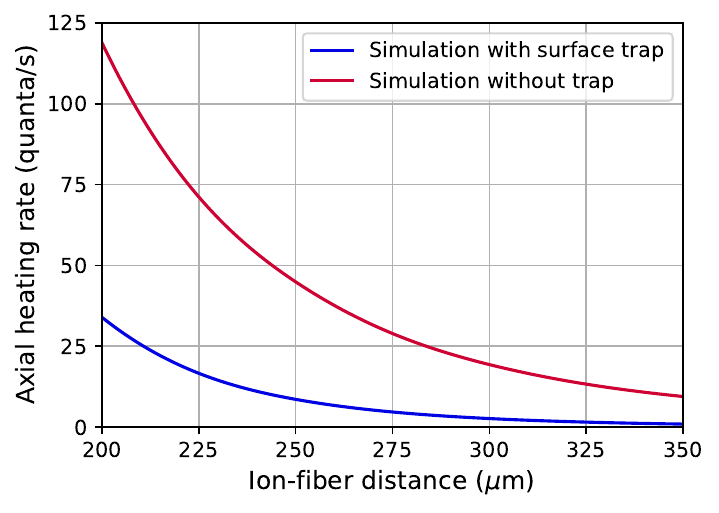}
\caption{Simulations of thermal-fluctuation-driven ion motional heating from a bare optical fiber when the ion is trapped above the grounded conductive surfaces of a surface electrode trap with the fiber attached directly to the trap, versus a geometry where the trap is absent and thus provides no reduction of the fields from the fiber at the ion position. Simulations are shown assuming 6.5\,K temperature, 1.5\,MHz axial frequency, and $\mathrm{\epsilon_r \tan{\delta} =  0.005}$, as discussed in the main text.}
\label{fig:heatingrate_trapcomparison}
\end{figure}

The low heating rates reported in the main text, relative to those from Reference~\cite{teller_heating_2021}, result from two main factors: (1) operation at cryogenic temperature reduces the thermal noise in the optical fiber, since that noise scales linearly with the temperature, and (2) the surface trap electrodes reduce the electric fields from the fiber at the ion position. Figure ~\ref{fig:heatingrate_trapcomparison} shows the thermal-noise-induced heating rate, simulated using finite-element analysis software as described in the main text. The results that include the surface trap in the simulation, with all trap electrodes grounded, are shown as a blue line. This simulation is also presented in the main paper and uses $\mathrm{\epsilon_r \tan{\delta} =  0.005}$, which is derived by fitting the axial heating rate data (see Figure\,3(a)). Removal of all nearby conducting surfaces from the simulation leads to a >4-fold increase in the heating rate (red line). Additionally, the distance scaling of the heating rate changes. Assuming that the heating rate has a power-law dependence $d^{-n}$, we find $n \approx 6.5$ when the surface trap is included, which changes to $n \approx 4.5$ when the surface trap is removed. This comparison highlights the benefit of using nearby conductive surfaces to reduce heating from a lossy dielectric.

Our work shows the viability of trapping ions near a dielectric material, for example in applications with miniaturized high-finesse optical cavities. High-reflectivity mirror coatings that comprise the cavity end mirrors commonly consist of $\mathrm{SiO_2}$/$\mathrm{Ta_2O_5}$ multilayers due to their low absorption loss at the parts-per-million level~\cite{rempe_measurement_1992}. We simulate the effect of a high-reflectivity mirror coating on the thermally induced heating rate using finite element analysis. The optical coating modeled consists of 46 layers, alternating between $\mathrm{Ta_2O_5}$ and $\mathrm{SiO_2}$, each with a thickness of 250\,nm. For $\mathrm{Ta_2O_5}$, we use the literature room-temperature values for the relative permittivity, with $\mathrm{\epsilon_{r,T} = 22}$ and $\mathrm{\tan{\delta_T} = 0.007}$~\cite{kim_tantala_2001, jain_tantala_2004}. For $\mathrm{SiO_2}$ and the optical fiber bulk, we use the fitted value $\mathrm{\epsilon_r \tan{\delta} =  0.005}$, as discussed in the main text. The simulation yields only a factor of $\approx 2$ increase in the heating rate when the optical coating is included. This indicates that low heating rates are achievable even with a high-finesse optical coating, provided that the cryogenic permittivity and loss tangent values of $\mathrm{Ta_2O_5}$, which to our knowledge are not currently documented at 6.5\,K, do not substantially deviate from room-temperature literature values.

\end{document}